\documentclass[twocolumn,letter]{jpsj3}
\usepackage{graphicx}% Include figure files
\usepackage{dcolumn}% Align table columns on decimal point
\usepackage{bm}% bold math
\usepackage{ulem}
\usepackage{braket}
\usepackage{amsmath}
\usepackage{color}
\usepackage{physics}
\usepackage[version=3]{mhchem}

\title{Engineering a skyrmion crystal in ferromagnetic/antiferromagnetic bilayers free from the DM interaction}

\author{Kazuki Okigami$^1$, Ryota Yambe$^1$, Satoru Hayami$^2$}
\inst{
$^1$Department of Applied Physics, The University of Tokyo, Tokyo 113-8656, Japan\\
$^2$Faculty of Science, Hokkaido University, Sapporo 060-0810, Japan 
}

\abst{
We theoretically propose a new stabilization mechanism of a skyrmion crystal (SkX) in a bilayer triangular lattice system consisting of the ferromagnetic and the antiferromagnetic layers. 
By performing variational calculations and Monte Carlo simulations in a complementary way, we find that a magnetic frustration between the ferromagnetic and antiferromagnetic layers is a source of a finite-$Q$ spiral state and the SkX in the strong interlayer coupling regime. 
We also show that the degree of frustration is related to the interlayer exchange interaction. 
The stronger interlayer coupling tends to make the effect of frustration larger, which results in the stabilization of the SkX.
The present results not only provide a way of engineering the SkX in the ferromagnetic/antiferromagnetic domain and heterostructure but also imply the possibility of the SkX based on interorbital frustration scenario. 
}

\begin{document}
\maketitle

The emergence of topologically nontrivial phenomena has drawn extensive interest as a new quantum state of condensed matter physics\cite{Nagaosa_RevModPhys.82.1539,Xiao_RevModPhys.82.1959,nagaosa2013topological,Baltz_RevModPhys.90.015005,batista2016frustration,smejkal2021anomalous}.
A magnetic skyrmion characterized by a topologically nontrivial swirling spin texture is a typical example to attract attention as an origin of emergent electromagnetic phenomena\cite{Bogdanov89,Bogdanov94,nagaosa2013topological,rossler2006spontaneous,Muhlbauer_2009skyrmion,yu2010real,yu2011near,Seki2012}.
When it is periodically aligned in crystals, namely a skyrmion crystal (SkX), a large physical response arising from the emergent electromagnetic fields through the spin Berry phase mechanism has been observed, such as the topological Hall and Nernst effects\cite{Neubauer_PhysRevLett.102.186602,Shiomi_PhysRevB.88.064409,Hamamoto_PhysRevB.92.115417,kurumaji2019skyrmion,Hirschberger_PhysRevLett.125.076602}.

Historically, the realization of the SkX has been revealed in the ferromagnetic (FM) systems under the noncentrosymmetric lattice structures\cite{rossler2006spontaneous,nagaosa2013topological,Yi_PhysRevB.80.054416,Binz_PhysRevLett.96.207202,Binz_PhysRevB.74.214408}, where the Dzyaloshinskii-Moriya (DM) interaction\cite{dzyaloshinsky1958thermodynamic,moriya1960anisotropic} plays an important role.
Then, it was shown that the instability toward the SkX in centrosymmetric magnets has been clarified in the triangular classical Heisenberg model in an external magnetic field by considering the competition between the FM and the antiferromagnetic (AFM) exchange interactions in addition to thermal fluctuations~\cite{Okubo_PhysRevLett.108.017206} and an easy-axis single-ion anisotropy~\cite{leonov2015multiply, Lin_PhysRevB.93.064430,Hayami_PhysRevB.93.184413}.
Furthermore, a different mechanism to stabilize the SkX has been proposed based on the long-range interaction represented as the Ruderman-Kittel-Kasuya-Yosida (RKKY) interaction in itinerant magnets\cite{Ruderman,Kasuya,Yosida1957}, where the itinerant nature of electrons that leads to an effective positive biquadratic interaction plays an important role\cite{Hayami_PhysRevB.95.224424,Ozawa_PhysRevLett.118.147205,hayami2021topological,wang2021skyrmion}.
More recently, the mechanism based on anisotropic exchange interactions depending on the lattice symmetry~\cite{yambe2022effective} or dipolar interaction has been shown under hexagonal\cite{Hayami_PhysRevB.103.054422,hayami2022skyrmion,Utesov_PhysRevB.105.054435}, tetragonal\cite{Christensen_PhysRevX.8.041022,hayami2020degeneracy,Hayami_PhysRevB.103.024439,Utesov_PhysRevB.103.064414,Wang_PhysRevB.103.104408,hayami2022multiple,Hayami_PhysRevB.105.104428,Hayami_PhysRevB.105.174437}, trigonal\cite{amoroso2020spontaneous,yambe2021skyrmion,amoroso2021tuning,hayami2022skyrmion}, and cubic systems\cite{hayami2021field}.
Simultaneously, the SkXs have been experimentally observed in centrosymmetric materials, such as \ce{Gd2PdSi3}\cite{Saha_PhysRevB.60.12162,kurumaji2019skyrmion,sampathkumaran2019report,Hirschberger_PhysRevB.101.220401,Kumar_PhysRevB.101.144440,Spachmann_PhysRevB.103.184424}, \ce{GdRu2Si2}\cite{khanh2020nanometric,Yasui2020imaging,khanh2022zoology}, Gd$_3$Ru$_4$Al$_{12}$\cite{hirschberger2019skyrmion,Hirschberger_10.1088/1367-2630/abdef9}, and EuAl$_4$\cite{Shang_PhysRevB.103.L020405,kaneko2021charge,zhu2022spin,takagi2022square}, while exhibiting the large Hall and Nernst effects owing to the short magnetic modulation periods compared to the purely DM-based mechanism.
Thus, the exploration of a new mechanism of the SkX is still an active research field in condensed matter physics.

In the present study, we propose another scenario to bring about the SkX by focusing on the role of the layer degree of freedom in a localized spin system. 
Specifically, we consider a bilayer triangular-lattice system consisting of the FM and AFM layers, which are ferromagnetically coupled.
For the bilayer spin model, we perform the variational calculations and Monte Carlo simulations.
As a result, we find that a magnetic frustration that arises from the layer competition gives rise to the SkX even without the DM interaction for the strong interlayer interaction.
We also construct the magnetic phase diagram while changing the temperature and the interlayer exchange interaction to show the stability region of the SkX.
Our result provides a guideline to engineer the SkX based on the layer degree of freedom.
We also discuss the relevance to another mechanism, the interorbital frustration\cite{Nomoto_PhysRevLett.125.117204}, which might be related to the origin of the SkX in Gd-based compounds.

\begin{figure}[t!]
\begin{center}
\includegraphics[width=1.0\hsize]{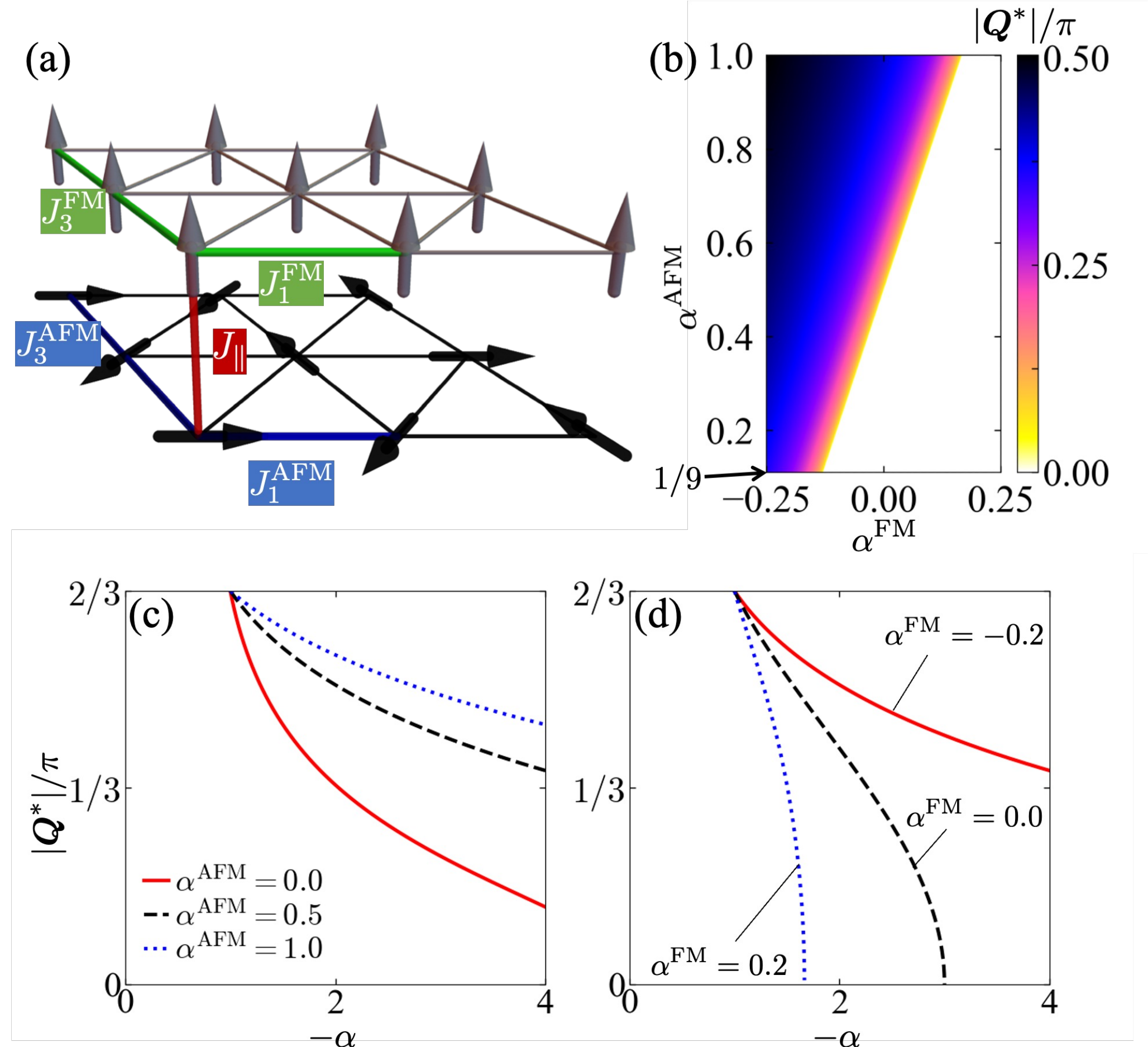} 
\caption{
\label{fig:model}
(a) Bilayer triangular lattice consisting of the FM layer hyperrefwith $J_1^{\rm FM}$ and $J_3^{\rm FM}$ and the AFM layer with $J_1^{\rm AFM}$ and $J_3^{\rm AFM}$. 
The FM and AFM layers are ferromagnetically coupled through $J_{\parallel}$. 
(b) The contour plot of the magnitude of the ordering vector $Q^* =|\vb*{Q}^*|$ in the plane of $\alpha^{\rm FM}$ and $\alpha^{\rm AFM}$ for $\alpha = J_1^{\rm FM}/ J_1^{\rm AFM}=-3$.
(c), (d) $\alpha$ dependence
of $Q^*$ for several values of (c) $\alpha^{\rm AFM}$ at $\alpha^{\rm FM}=-0.2$ and those of (d) $\alpha^{\rm FM}$ at $\alpha^{\rm AFM}=0.5$.
}
\end{center}
\end{figure}

Let us start by showing a bilayer triangular-lattice model in Fig.~\ref{fig:model}(a), whose Hamiltonian is given by
\begin{align}
    \label{eq:Model}
    \mathcal{H} &= \sum_{\eta={\rm FM,AFM}} \mathcal{H}_{\eta} + \mathcal{H}_{\parallel}, \\
    \label{eq:Heta}
    \mathcal{H}_{\eta} &=  
        -J_1^{\eta} \sum_{\langle{i,j}\rangle} \vb*{S}_i^{\eta} \cdot \vb*{S}_j^{\eta}
        - J_3^{\eta} \sum_{\langle\langle\langle{i,j}\rangle\rangle\rangle} \vb*{S}_i^{\eta} \cdot \vb*{S}_j^{\eta}, \\
    \label{eq:Hparallel}
    \mathcal{H}_{\parallel} &=    
        -J_{\parallel} \sum_i \vb*{S}_i^{\rm{FM}} \cdot \vb*{S}_i^{\rm{AFM}},
\end{align}
where $\vb*{S}_i^{\eta}$ is a classical localized spin at site $i$ on the layer $\eta\ (={\rm FM,\ AFM})$ with $|\vb*{S}_i^{\eta}|=1$.
The total Hamiltonian $\mathcal{H}$ in Eq.~(\ref{eq:Model}) is composed of the intralayer Hamiltonian $\mathcal{H}_{\eta}$ for $\eta=$ FM, AFM in Eq.~(\ref{eq:Heta}) and the interlayer Hamiltonian $\mathcal{H}_{\parallel}$ in Eq.~(\ref{eq:Hparallel}).
In Eq.~(\ref{eq:Heta}), $J_1^{\eta}$ ($J_3^{\eta}$) stands for the (third) nearest-neighbor coupling constant for layer $\eta$; $\sum_{\langle{i,j}\rangle}$ ($\sum_{\langle\langle\langle{i,j}\rangle\rangle\rangle}$) denotes the sum of the (third) nearest-neighbor pairs. 
Although we consider the $J_1$-$J_3$ model, the following results are applicable to a different set of the coupling constants including the other $\iota$th nearest-neighbor coupling constants by appropriately choosing the model parameters as discussed below.
In Eq.~(\ref{eq:Hparallel}), $J_{\parallel}$ represents the interlayer coupling constant.
 
Hereafter, we set $J_{1}^{\rm AFM}=-1$ as the energy unit.
The other interaction parameters in Eq.~(\ref{eq:Heta}), $J_1^{\rm FM}$, $J_3^{\rm FM}$, and $J_{3}^{\rm AFM}$, are set so that the ground-state spin configuration in the layer $\eta={\rm FM}$ and AFM becomes the FM and 120$^{\circ}$ AFM ordering, respectively, for $J_{\parallel}=0$ as follows: $J_1^{\rm FM} >0$ and $\alpha^{\rm FM} = J_3^{\rm FM}/J_1^{\rm FM}>-1/4$ for the FM layer and $\alpha^{\rm AFM} = J_3^{\rm AFM}/J_1^{\rm AFM}>1/9$ for the AFM layer.
In addition, we introduce the ratio of the interaction between the FM and AFM layers $\alpha=J_1^{\rm FM}/J_1^{\rm AFM}$.
The remaining parameter $J_{\parallel}$ is set to be FM as $J_{\parallel}>0$.

We first focus on the strong coupling limit of the FM interlayer interaction $J_{\parallel} \to \infty$, where the spins for the FM layer are parallel to those for the AFM layer, i.e., $\vb*{S}_i = \vb*{S}_i^{\rm{FM}} = \vb*{S}_i^{\rm{AFM}}$. 
In such a situation, the bilayer Hamiltonian in Eq.~(\ref{eq:Model}) is mapped onto the effective single-layer Hamiltonian written as
\begin{align}
    \begin{split}
        \mathcal{H}^{\rm eff} &= 
            -J_1^{\rm eff} \sum_{\langle{i,j}\rangle} \vb*{S}_i \cdot \vb*{S}_j
            - J_3^{\rm eff} \sum_{\langle\langle\langle{i,j}\rangle\rangle\rangle} \vb*{S}_i \cdot \vb*{S}_j,
        \label{eq:effModel}
    \end{split}
\end{align}
where $J_1^{\rm eff}=J_1^{\rm FM}+J_1^{\rm AFM}$ and $J_3^{\rm eff}=J_3^{\rm FM}+J_3^{\rm AFM}$. 
Then, we focus on the situation where the ground state becomes the incommensurate spiral state by taking $J_1^{\rm eff} >0 (\alpha<0)$ and $J_3^{\rm eff}/J_1^{\rm eff} = (\alpha \alpha^{\rm FM}+\alpha^{\rm AFM})/(\alpha+1) <-1/4$. 
The magnitude of the ordering vector $\vb*{Q}^*$, $Q^* = |\vb*{Q}^*|$, is given by 
\begin{align}
    Q^* = 2 \cos^{-1} 
    \left[\frac{1}{4} \left(1+\sqrt{1-\frac{2(\alpha+1)}{\alpha\alpha^{\rm FM}+\alpha^{\rm AFM}}} \right) \right],
\end{align}
where the lattice constant is set as unity. 
$Q^*$ is also rewritten as $Q^* = 2 \cos^{-1} \left[(1+\sqrt{1-2J_1^{\rm eff}/J_3^{\rm eff}} ) / 4 \right]$\cite{tamura2011first}. 
Owing to sixfold rotational symmetry of the lattice, there are six equivalent ordering vectors; $\pm \vb*{Q}_1^*=\pm Q^* \hat{x}$, $\pm \vb*{Q}_2^*=\pm Q^* \left(-\frac{1}{2}\hat{x} + \frac{\sqrt{3}}{2}\hat{y} \right)$, $\pm \vb*{Q}_3^*=\pm Q^* \left(-\frac{1}{2}\hat{x} - \frac{\sqrt{3}}{2}\hat{y} \right)$ with unit vectors $\hat{x}$ and $\hat{y}$ along the $x$ and $y$ directions, respectively.

Figure~\ref{fig:model}(b) shows the contour plot of $Q^*$ against $\alpha^{\rm FM}$ and $\alpha^{\rm AFM}$ at $\alpha=-3$, where $Q^*$ continuously changes while changing $\alpha^{\rm FM}$ and $\alpha^{\rm AFM}$.
The result in Fig.~\ref{fig:model}(b) indicates that small $\alpha^{\rm FM}$ and large $\alpha^{\rm AFM}$ tend to lead to the instability toward the spiral state with the incommensurate ordering vector; the large ordering vector (short magnetic period) is obtained for 
small $\alpha^{\rm FM}$ ($J_3^{\rm FM} <0 $) and 
large $\alpha^{\rm AFM}$ ($J_3^{\rm AFM} <0$).
It is noted that the spiral state is replaced by the FM state with $Q^*=0$ when ignoring the further-neighbor interaction $\alpha^{\rm FM}=\alpha^{\rm AFM}=0$. 
A similar tendency is found for other $\alpha$, as shown in Figs.~\ref{fig:model}(c) and \ref{fig:model}(d).
Hereafter, we set $Q^*\simeq0.42\pi\ (J_3^{\rm eff}/J_1^{\rm eff}=0.55)$ by choosing $J_1^{\rm FM}=3$, $J_3^{\rm FM}=-0.6$, and $J_3^{\rm AFM}=-0.5$, which corresponds to $\alpha=-3$, $\alpha_{\rm FM}=-0.2$, and $\alpha_{\rm AFM}=0.5$.

\begin{figure}[t!]
\begin{center}
\includegraphics[width=1.0\hsize]{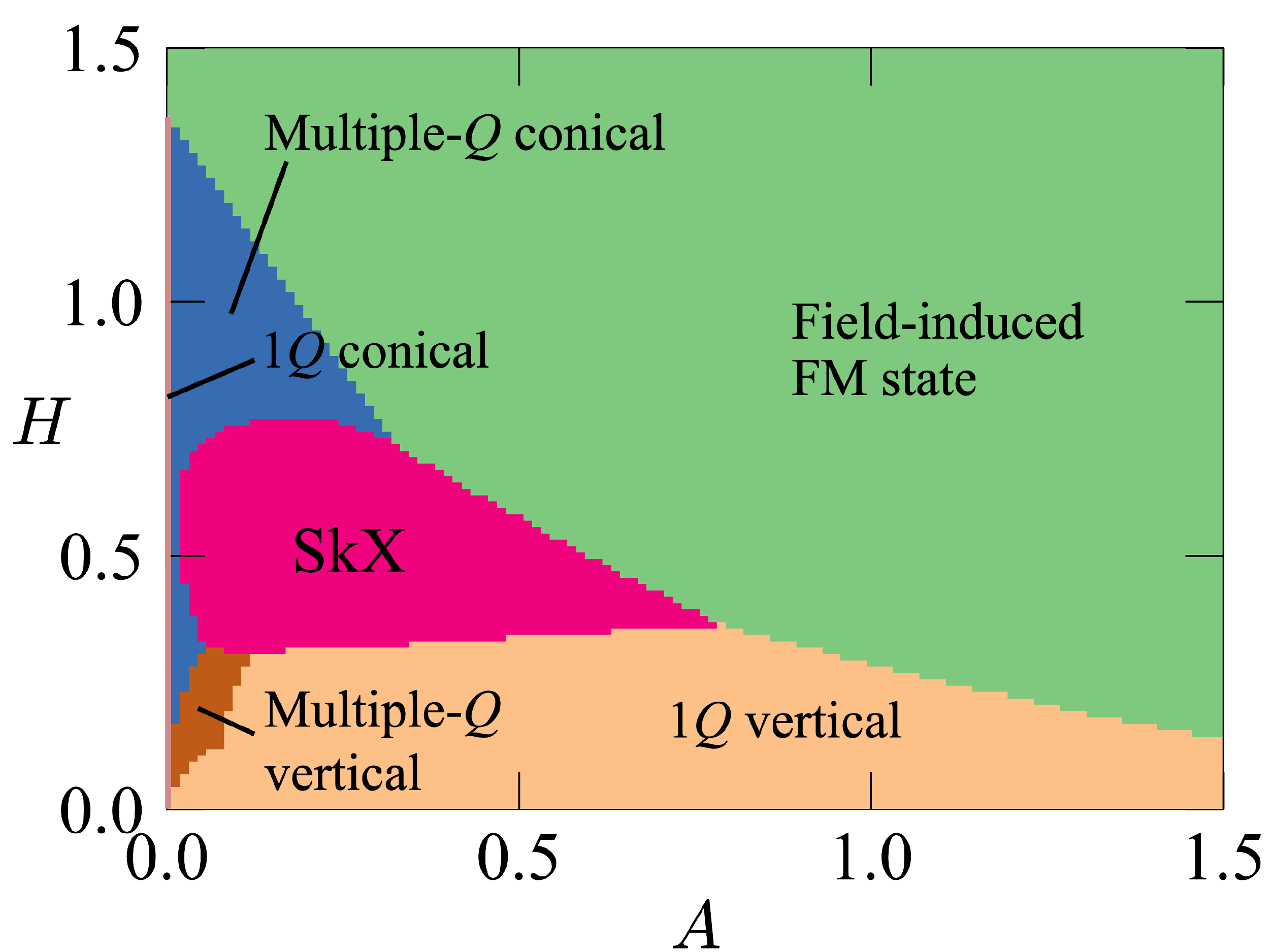} 
\caption{
\label{fig:PD}
Ground-state phase diagram for the model $\mathcal{H}+\mathcal{H}_{\rm loc}$ in the strong interlayer coupling regime, $J_{\parallel}=500$, by variational calculations on the bilayer triangular lattice for $J_1^{\rm FM}=3$, $J_3^{\rm FM}=-0.6$, $J_1^{\rm AFM}=-1$, $J_3^{\rm AFM}=-0.5$. 
The SkX represents the skyrmion crystal.
}
\end{center}
\end{figure}

As the model in Eq.~(\ref{eq:effModel}) is the same as the single-layer frustrated spin model, the SkX can appear as the ground state by taking into account the magnetic field ($H$) along the $z$ direction and the easy-axis single-ion anisotropy ($A$), whose Hamiltonian is given by $\mathcal{H}_{\rm loc} = -\sum_{\eta,i} \left[H S_{i,z}^{\eta} + A  (S_{i,z}^{\eta})^2 \right]$ with $A>0$. 
For the Hamiltonian $\mathcal{H}+\mathcal{H}_{\rm loc}$
with large $J_{\parallel}=500$, we perform the variational calculations by considering the following six spin configurations: 

\begin{itemize}
\item[(a)]
The SkX with 
\begin{align}
    \begin{split}
        S_{i,x} &= a_1 \left(-\frac{\sqrt{3}}{2} \sin \mathcal{Q}_{i,2} + \frac{\sqrt{3}}{2} \sin \mathcal{Q}_{i,3} \right), \\
        S_{i,y} &= a_1 \left(\sin \mathcal{Q}_{i,1} - \frac{1}{2} \sin \mathcal{Q}_{i,2} - \frac{1}{2} \sin \mathcal{Q}_{i,3} \right), \\
        S_{i,z} &= -a_2 \left(\cos \mathcal{Q}_{i,1} + \cos \mathcal{Q}_{i,2} + \cos \mathcal{Q}_{i,3} \right) + \Tilde{m},
    \end{split}
\end{align}
where $\mathcal{Q}_{i,\nu} = \vb*{Q}_{\nu}^* \cdot \vb*{r}_i$; $\vb*{r}_i$ is the position vector at site $i$.
\item[(b)]
The single-$Q$ conical spiral with $\vb*{S}_i = (\sin \theta \cos \mathcal{Q}_{i,\nu},\ \sin\theta \sin \mathcal{Q}_{i,\nu},\ \cos\theta)$. 
\item[(c)]
The multiple-$Q$ conical spiral with $\vb*{S}_{i}
 = (a_1 \cos \mathcal{Q}_{i,1} + a_2 \cos \mathcal{Q}_{i,2},\ -a_1 \sin \mathcal{Q}_{i,1} + a_2 \sin \mathcal{Q}_{i,2},\ a_3 \cos \mathcal{Q}_{i,3} + \Tilde{m})$.
\item[(d)]
The single-$Q$ vertical spiral with $\vb*{S}_i = (0,\ a_1 \sin \mathcal{Q}_{i,1},\ a_2 \cos \mathcal{Q}_{i,1} +\Tilde{m})$.
\item[(e)]
The multiple-$Q$ vertical with $\vb*{S}_i = (a_1 \cos (\mathcal{Q}_{i,2} + \phi) - a_1 \cos (\mathcal{Q}_{i,3} - \phi),\ 
-a_2 \sin \mathcal{Q}_{i,1},\ 
a_2 \cos \mathcal{Q}_{i,1} + \Tilde{m})$.
\item[(f)]
The FM state with $\vb*{S}_i=(0,\ 0,\ 1)$.
\end{itemize}
The spin length in each state is normalized as $|\vb*{S}_i|=1$.
We take $\vb*{S}_i = \vb*{S}_i^{\rm{FM}} = \vb*{S}_i^{\rm{AFM}}$ due to the strong interlayer coupling $J_{\parallel}=500$. 
$a_{\nu}$, $\Tilde{m}$, $\theta$ and $\phi$ are the variational parameters.
We do not consider the relative phase among the constituent waves in the SkX for simplicity~\cite{Hayami_PhysRevResearch.3.043158,Shimizu_PhysRevB.105.224405}.
We set the total sites $N=2\times L^2$ with $L=48$, where the states with $Q^*=5 \pi /12$ become the lowest-energy state.

Figure~\ref{fig:PD} shows the $H$-$A$ phase diagram at the zero temperature $T=0$. 
The single-$Q$ conical spiral state stabilized at $A=0$ turns into the multiple-$Q$ conical state for infinitesimal $A$~\cite{Hayami_PhysRevB.93.184413}. 
While further increasing $A$, the SkX, the multiple-$Q$ vertical, and the single-$Q$ vertical spiral states are stabilized. 
In particular, we find that the SkX with a short period is robustly stabilized in the bilayer system for intermediate $H$ and $A$.
Although the emergence of the SkXs has been revealed in the multi-layer systems based on the interfacial DM interaction\cite{jiang2015blowing,woo2016observation,dohi2022thin} and the layer-dependent (staggered) DM interaction\cite{Hayami_PhysRevB.105.014408,lin2021skyrmion,Hayami_PhysRevB.105.184426,Hayami_2022}, the present mechanism is closely related to that based on the frustrated exchange interaction instead of the DM interaction.  
Indeed, the result is qualitatively similar to that for the frustrated spin model on a single-layer triangular lattice~\cite{leonov2015multiply,Lin_PhysRevB.93.064430}, since the present bilayer Hamiltonian reduces to the single-layer Hamiltonian for large $J_{\parallel}$, as discussed above. 

\begin{figure}[t!]
\begin{center}
\includegraphics[width=1.0\hsize]{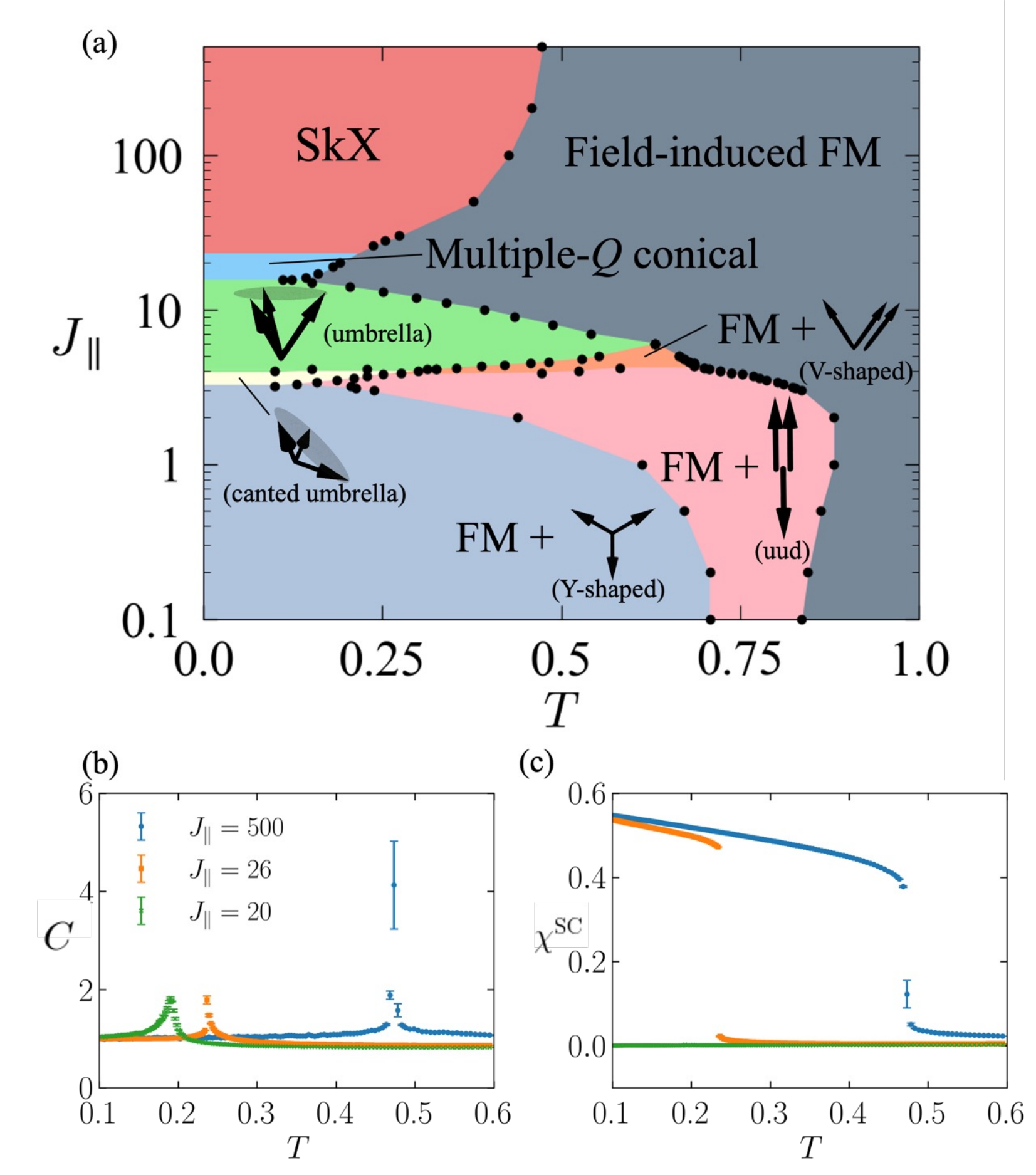} 
\caption{
\label{fig:MC}
(a) $J_{\parallel}$-$T$ phase diagram obtained by replica exchange Monte Carlo simulations at $A=0.25$ and $H=0.5$.
The $y$-axis is in logarithmic scale.
The other parameters are the same as in Fig.~\ref{fig:PD}.
(b), (c) Temperature dependence of (b) the specific-heat $C$ and (c) the scalar chirality $\chi^{\rm SC}$ for $J_{\parallel}=20,26,500$.
}
\end{center}
\end{figure}

Next, we examine the stability of the SkX at finite $T$ while decreasing $J_{\parallel}$. 
We fix $A=0.25$ and $H=0.5$ and perform an unbiased Monte Carlo (MC) simulation based on the Metropolis algorithm. 
We also combine the replica-exchange method\cite{hukushima1996exchange}. 
In the simulations, we adopt two update schemes to reduce autocorrelations: One is the single-spin update of $\vb*{S}_i^{\eta}$ and the other is the two-spin update of $\vb*{S}_i^{{\rm FM}}$ and $\vb*{S}_i^{{\rm AFM}}$ while keeping the inner product $\vb*{S}_i^{{\rm FM}} \cdot \vb*{S}_i^{{\rm AFM}}$. 
Each run has a total of $10^7$-$10^8$ MC sweeps to reach equilibrium and $10^7$-$10^8$ MC sweeps to calculate the average of the physical variables.
The phase diagram is constructed for the system sizes with $L=36$, $48$, and $60$ under the periodic boundary conditions. 
As the system-size dependence was small, we show the result at $L=48$ below.

The obtained $J_{\parallel}$-$T$ phase diagram is presented in Fig.~\ref{fig:MC}(a).
The transition points are identified from the specific heat and the magnetic susceptibility.
While decreasing $J_{\parallel}$, one finds that the SkX is stable for $26 \lesssim  J_{\parallel} \leq \infty$, which means that the SkX appears only for large $J_{\parallel}$. 
This state is also stabilized at finite temperatures, where the transition temperature becomes larger for larger $J_{\parallel}$. 
The transition from the field-induced FM (paramagnetic) state to the SkX is clearly identified in the peak structure of the specific heat $C$ in Fig.~\ref{fig:MC}(b) and the jump of the total scalar chirality $\chi^{\rm SC}$ in Fig.~\ref{fig:MC}(c) in the case of $J_{\parallel}=500$ and $26$; the first-order phase transition occurs similar to the single-layer case\cite{Lin_PhysRevB.93.064430,Hayami_PhysRevB.93.184413}.  
Here, $\chi^{\rm SC}=\sqrt{\expval{\frac{1}{4N} (\sum_{i,\eta} \chi_{i,\eta}^{\rm SC})^2}}$ with $\chi_{i,\eta}^{\rm SC}=\vb*{S}_{i_1}^{\eta} \cdot (\vb*{S}_{i_2}^{\eta} \times \vb*{S}_{i_3}^{\eta})$, where $i_1,\ i_2,\ i_3$ are aligned anticlockwise on a single triangle $i$ and $\langle \cdots \rangle$ denotes the thermodynamic average.

\begin{figure}[t!]
\begin{center}
\includegraphics[width=1.0\hsize]{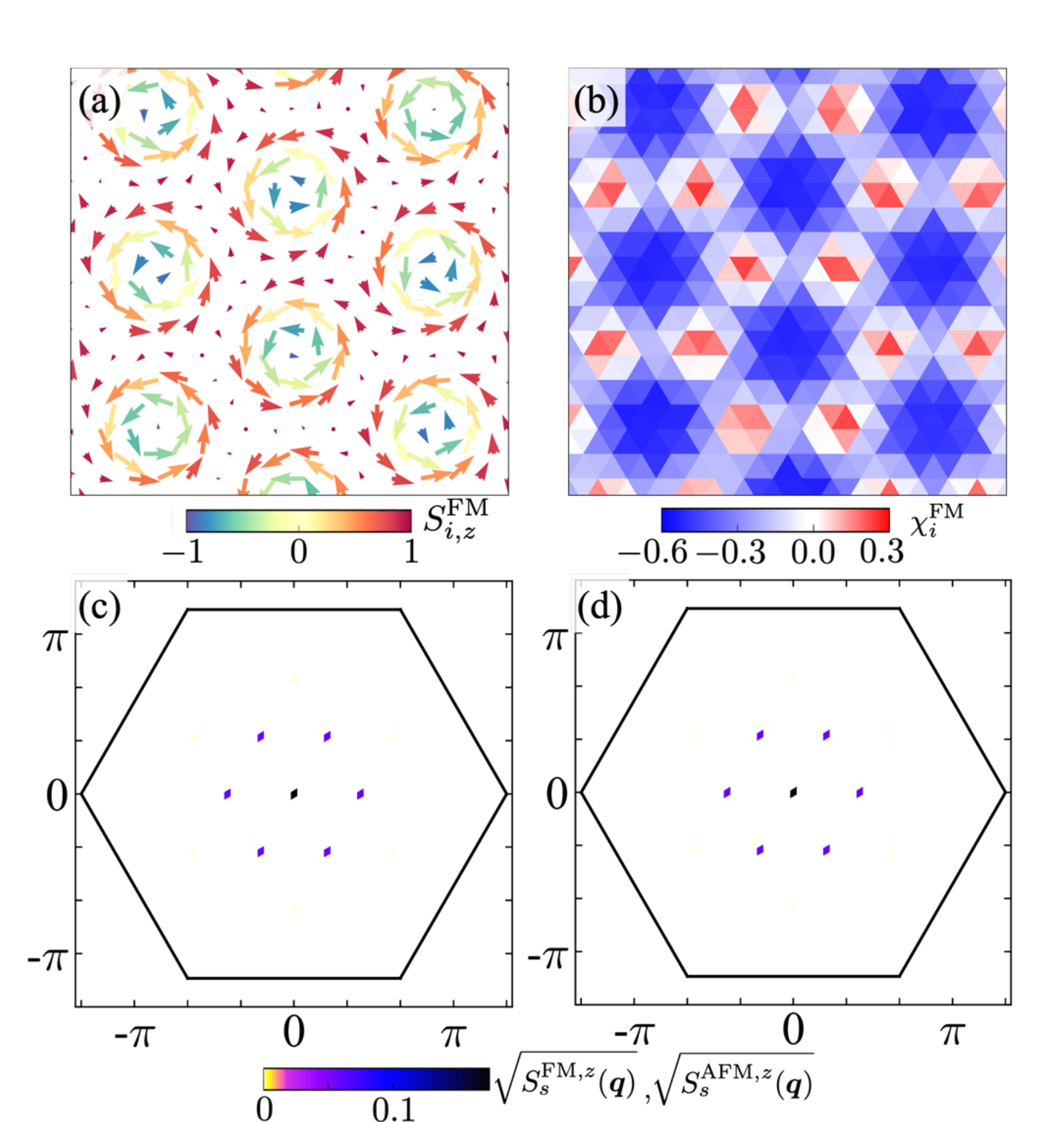} 
\caption{
\label{fig:snapshots}
(a) Snapshot of the spin configuration in the SkX for $J_{\parallel}=-500$, $A=0.25$, and $H=0.5$ at $T=0.1$.
The arrow shows the inplane spin component, while the color shows the out-of-plane spin component.
Spin components are averaged over 100 times. 
(b) Contour plot of the local scalar chirality obtained from the spin configuration in (a).
(c) $\sqrt{S_s^{{\rm FM},z} (\vb*{q})}$ and (d) $\sqrt{S_s^{{\rm AFM},z} (\vb*{q})}$ in the SkX; the black lines stand for the first Brillouin zones.
}
\end{center}
\end{figure}

In the SkX phase, the spin configuration consists of the periodic array of the skyrmions, as shown in the case of the FM layer in Fig.~\ref{fig:snapshots}(a).
The spin texture shows a nonzero scalar chirality in Fig.~\ref{fig:snapshots}(b), which leads to a quantized skyrmion number of $-1$.
Since the spin configuration on the AFM layer is almost the same as that on the FM layer owing to the strong interlayer coupling, there is no cancellation of the scalar chirality (skyrmion number).
In momentum space, the out-of-plane spin structure factor $S_s^{\eta,z} (\vb*{q}) = (1/N) \sum_{jl} \langle S_{j,z}^{\eta} S_{l,z}^{\eta} \rangle e^{i \vb*{q} \cdot (\vb*{r}_j - \vb*{r}_l)}$ exhibits the six peaks at $\pm \vb*{Q}^*_{\nu}$ [we define $\bm{Q}^*_1 \equiv \bm{Q}_{\rm SK}$ ($|\bm{Q}_{\rm SK}|=5\pi/12$)] in addition to $\vb*{q}=\vb*{0}$ for both FM and AFM layers, as shown in Figs.~\ref{fig:snapshots}(c) and \ref{fig:snapshots}(d), respectively. 

While further decreasing $J_{\parallel}$, the SkX is replaced by the multiple-$Q$ conical state without the scalar chirality, as shown in Figs.~\ref{fig:MC}(b) and \ref{fig:MC}(c).
The spin configurations are almost the same for the FM and AFM layers similar to the SkX phase. Meanwhile, the magnitude of the ordering vector is different from that of the SkX; the ordering vector is given as $\bm{Q}_{\rm MQ}$ ($|\bm{Q}_{\rm MQ}|=\pi/2$) in this state. 

\begin{figure}[t!]
\begin{center}
\includegraphics[width=1.0\hsize]{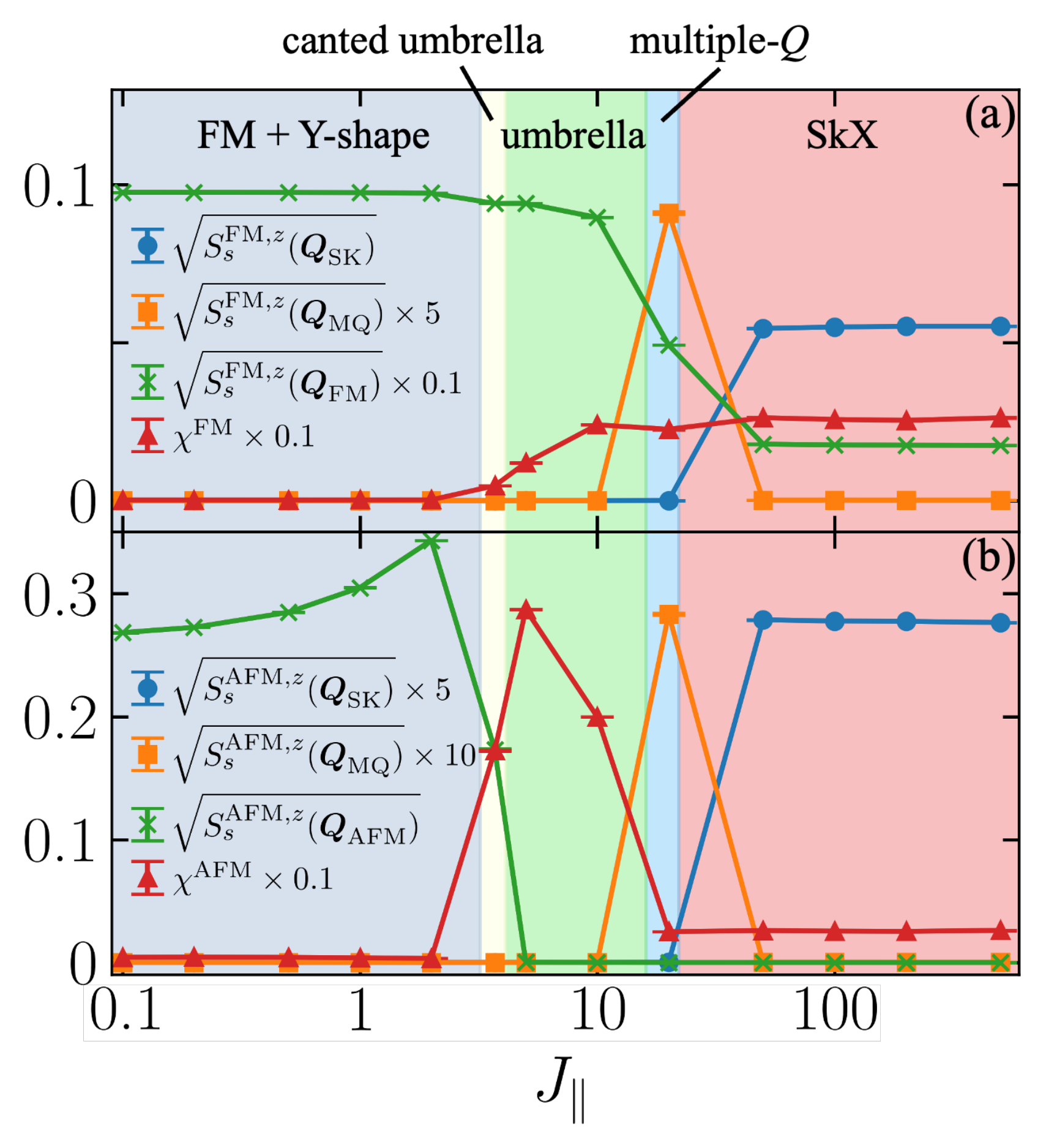} 
\caption{
\label{fig:sfchi}
$J_{\parallel}$ dependence of the spin structure factors $\sqrt{S_s^{\eta,z} (\vb*{q})}$ at $\bm{Q}_{\rm SK} (5/12\pi,0)$, $\bm{Q}_{\rm MQ} =(\pi/2,0)$, $\bm{Q}_{\rm FM}=(0,0)$, $\bm{Q}_{\rm AFM}=(4/3\pi,0)$ and the vector chiralities for (a) the FM layer and (b) the AFM layer at $T=0.1$.}
\end{center}
\end{figure}

When $3 \lesssim J_{\parallel} \lesssim 15$, both the FM and AFM layers show the three-sublattice (canted) umbrella spin texture but they are different from each other. 
For example, $\vb*{S}_i^{\rm FM} \cdot \vb*{S}_i^{\rm AFM} \sim 0.8$ at $J_{\parallel}=10$ and $T=0.1$ so that the spins of the FM layer have a more $z$ component than those of the AFM layer.
For $J_{\parallel} \lesssim 3$, the FM and AFM layers show the different spin configurations with different $\bm{q}$-peak structures; the FM layer shows the FM spin configuration with the peak at $\bm{Q}_{\rm FM}=\bm{0}$, while the AFM layer shows the three-sublattice AFM spin configurations with the peak at $\bm{Q}_{\rm AFM}=(4\pi/3,0)$. 
Especially, the three-sublattice AFM spin configurations on the AFM layer depend on $J_{\parallel}$ and $T$, whose schematic pictures are denoted in Fig.~\ref{fig:MC}(a). 
Each spin configuration is distinguished by the spin structure factor in addition to the vector chirality $\chi^{\eta}=\sum_i \chi_{i}^{\eta}$ with $\chi_{i}^{\eta} = (\vb*{S}_{i_1}^{\eta} \times \vb*{S}_{i_2}^{\eta} + \vb*{S}_{i_2}^{\eta} \times \vb*{S}_{i_3}^{\eta} + \vb*{S}_{i_3}^{\eta} \times \vb*{S}_{i_1}^{\eta})_z$ on the triangle consisting of the sites $i_1$, $i_2$, and $i_3$. 
We show the $J_{\parallel}$ dependence of these quantities for the FM and AFM layers at $T=0.1$ in Figs.~\ref{fig:sfchi}(a) and \ref{fig:sfchi}(b), respectively. 

In summary, we have investigated a bilayer triangular-lattice system composed of the FM and the AFM layers, where each layer only has its own trivial ground state when there is no interlayer interaction.
By performing variational calculations and MC simulations, we obtained the instability toward the incommensurate spiral state and the SkX with the short magnetic modulation periods by taking into account the strong FM interlayer interaction.
The results provide a new stabilization mechanism of the short-period SkX based on the layer degree of freedom even without the DM interaction, which will stimulate further exploration of the SkXs in the domain structures and heterostructures.

Finally, let us discuss the relevance of the present mechanism to that based on interorbital frustration that has been recently proposed for the origin of the incommensurate spiral ordering in the Gd-based compounds, Gd$_2$PdSi$_3$\cite{Saha_PhysRevB.60.12162,kurumaji2019skyrmion,sampathkumaran2019report,Hirschberger_PhysRevB.101.220401,Kumar_PhysRevB.101.144440,Spachmann_PhysRevB.103.184424} and GdRu$_2$Si$_2$\cite{khanh2020nanometric,Yasui2020imaging,khanh2022zoology}. 
There, the importance of the magnetic frustration arising from the competition between the Gd-5{\it d} FM interaction and the Gd-4{\it f} AFM interaction has been implied based on the {\it ab-initio} calculations\cite{Nomoto_PhysRevLett.125.117204}. 
Such a situation can be roughly mapped onto the present bilayer model when supposing that the FM (AFM) layer mimics the Gd-5{\it d} (Gd-4{\it f}) electrons, where the interlayer coupling corresponds to the {\it d}-{\it f} exchange interaction.
Our result indicates that the {\it d}-{\it f} exchange interaction should be much larger than the {\it d}-{\it d} and {\it f}-{\it f} ones to induce the instability toward the SkX. 
Furthermore, magnetic anisotropy like the single-ion anisotropy is required to realize the ground-state SkX.
It is desired to quantitatively evaluate these interactions and anisotropy by the {\it ab-initio} calculations to investigate the possibility of the SkX based on the interorbital frustration mechanism in the Gd-based compounds, which will be left for a future intriguing problem. 

\begin{acknowledgments}
This research was supported by JSPS KAKENHI Grants Numbers JP21H01037, JP22H04468, JP22H00101, JP22H01183 and by JST PRESTO (JPMJPR20L8). 
R.Y. was supported by Forefront Physics and Mathematics Program to Drive Transformation (FoPM).
Parts of the numerical calculations were performed in the supercomputing systems in ISSP, the University of Tokyo.
\end{acknowledgments}

\bibliographystyle{JPSJ}
\bibliography{main_final.bbl}

\end{document}